\documentclass{llncs}

\usepackage{makeidx}  

\usepackage{algorithm,algorithmic,amsmath,amssymb,color,comment,courier,dcolumn,float,graphicx,helvet,latexsym,subfigure,tikz,times,url}

\makeindex

\begin{document}

\frontmatter          
\pagestyle{headings}  

\title{Topic Lifecycle on Social Networks: Analyzing the Effects of Semantic Continuity and Social Communities}
\titlerunning{Topic Lifecycle on Social Networks: Semantic Continuity and Social Communities}


\author{Kuntal Dey, Saroj Kaushik, Kritika Garg, Ritvik Shrivastava}
\institute{
Kuntal Dey, IBM Research, New Delhi, India. \email{kuntadey@in.ibm.com}
\and Saroj Kaushik, IIT, Delhi, India. \email{saroj@cse.iitd.ac.in}
\and Kritika Garg, Ch. Brahm Prakash GEC, New Delhi, India. \email{kgarg.kritika@gmail.com}
\and Ritvik Shrivastava, NSIT, New Delhi, India. \email{ritviks.it@nsit.net.in}
}

\maketitle

\abstract{
Topic lifecycle analysis on Twitter, a branch of study that investigates Twitter topics from their birth through lifecycle to death, has gained immense mainstream research popularity.
In the literature, topics are often treated as one of (a) hashtags (independent from other hashtags), (b) a burst of keywords in a short time span or (c) a latent concept space captured by advanced text analysis methodologies, such as Latent Dirichlet Allocation (LDA).
The first two approaches are not capable of recognizing topics where different users use different hashtags to express the same concept (semantically related), while the third approach misses out the user's explicit intent expressed via hashtags.
In our work, we use a word embedding based approach to cluster different hashtags together, and the temporal concurrency of the hashtag usages, thus forming topics (a semantically and temporally related group of hashtags).
We present a novel analysis of topic lifecycles with respect to communities.
We characterize the participation of social communities in the topic clusters, and analyze the lifecycle of topic clusters with respect to such participation.
We derive first-of-its-kind novel insights with respect to the complex evolution of topics over communities and time: temporal morphing of topics over hashtags within communities, how the hashtags die in some communities but morph into some other hashtags in some other communities (that, it is a community-level phenomenon), and how specific communities adopt to specific hashtags.
Our work is fundamental in the space of topic lifecycle modeling and understanding in communities: it redefines our understanding of topic lifecycles and shows that the social boundaries of topic lifecycles are deeply ingrained with community behavior.
}

\section{Introduction}
\label{sec:intro}

Twitter has been a key social network platform for diffusion of information via user interactions.
Several research works have been carried out, that analyze the user-generated content, to identify the characteristics of information diffusion.
One core research area has focused on the topics present in user-generated content, either via hashtag analysis or sophisticated text-analytics driven derivations.
And based upon that, research has further focused on identifying the topics of user interest, and understanding the lifecycle of these topics - how these topics emerge, how they spread over the social network successfully (or not) and proliferate across several users, and eventually how they subside over time.

Some works in the literature have attempted to investigate lifecycles of topics.
In a pioneering work, Ardon {\it et al.} \cite{ardon2013spatio} investigated the shape and rate of adoption of topics among social users, where they treated hashtags as topics.
They observed that, topics (hashtags) have a five-phase lifecycle, peaking in the middle phases.
They presented a detailed study of the social graphs associated with the topics, such as the degree distributions, the presence (and essence) of giant components, and geographical distributions.

Other works have also attempted to understand the lifecycle of topics; however, they have focused on the linguistic aspects more than the social aspects, and have treated the topic lifetime problem (how long a topic lasts, without focusing on {\it socially with whom}) in the form of a hashtag disambiguation problem.
In an early work, Yang and Leskovec \cite{yang2011patterns} detected similar distributions of usage of given Twitter hashtags in form of temporal usage shapes, using K-Spectral Centroid (KSC) clustering.
However, this work did not investigate (a) the temporal overlap of different hashtags - whether or not a given pair of hashtags occurs at similar times, (b) the semantic concept space addressed by the corresponding tweets - if two different hashtags originate from tweets with the same meaning then it goes uncaptured, and (c) the social angle was completely missing too.
In a recent work, Stilo and Velardi \cite{stilo2017hashtag} proposed SAX, a temporal sense clustering algorithm based on the hypothesis that semantically related hashtags have similar and synchronous usage patterns.
Thus, SAX overcomes a key shortcoming of KSC by considering the temporal overlap of different hashtags.
However, it still does not account for the social angle; and in addition, does not attempt to consider the semantic space overlap across hashtags, which in turn leads to clustering of topically unrelated tweets also.
Further, none of these approaches attempt to understand the morphing of topics and whether intricate social community interaction dynamics are associated with any such morphing.

On the contrary, we believe that, social communities (that are formed purely based upon familiarity structures), and the intricacy of interactions of users, are the core determinants of topic lifecycle - how topics are born, how they spread, and how they die and morph.
However, we note that, in order to understand topics in the true sense, one needs to first acknowledge that, (a) in reality topics spread over and beyond a single hashtag: {\it \#federer} and {\it \#rogerfederer} are the same topics really, and (b) considering the latent semantic concept space of tweets is insufficient to account for the user's intent unless the hashtag is also considered: ``I love him'' is not the same as ``I love him \#Obama'' - the former is probably a simple expression of personal love while the later is clearly a political expression.
Hence, we propose a novel technique to bring related hashtags together by clustering as a combination of the semantic space ({\it \#NFL} is National Football League for sports but National Fertilizers Limited for agriculture), hashtags and the time of expression (\#USOpen is ``obviously'' the golf tag during the golf time but the tennis tag during the tennis time).
We hypothesize that, hashtags, and topics derived using the hashtags, bear the following characteristics.

\begin{itemize}
\item {\bf Hypothesis 1 - Conceptually related hashtags overlap semantically and temporally:} Different users use different hashtags at the same time for the same topic, that are semantically related and temporally overlapping. That is, one user would use {\it \#wimbledon} while another would use {\it \#bigW}, but their content would semantically (conceptually) overlap, and the usage would be temporally around similar (overlapping) times too.
\item {\bf Hypothesis 2 - Hashtags associate with communities at a given time:} Hashtag usages are community-level characteristics rather than individual-level. Individuals mostly tend to use the same hashtag that their community would use, for a given topic, at a given time. That is, if two users $u_1$ and $u_2$ belong to the same community, then they both are likely to use {\it \#federer} instead of one using {\it \#federer} and the other using {\it \#rogerfederer}.
\item {\bf Hypothesis 3 - Hashtags are independently used across communities:} Inter-community independence of hashtag usage is an inherent property of social networks. That is, for the same topic, at the same time, while one community would use one hashtag, another community would use another hashtag. That is, community $C_1$ as a whole would tend to use the term {\it \#federer} while community $C_2$ as a whole would tend to use the term {\it \#rogerfederer}.
\item {\bf Hypothesis 4 - Hashtags evolve independently (atomically) within communities:} Evolution and lifecycle of hashtags (and topics) are community specific. The global (overall) lifecycle of a given topic can be derived as an aggregation of the lifecycle of topics within individual communities, along an overall span of time. For example, in a given span of 7 days, community $C_1$ would use the hashtag {\it \#federer} for the first 2 days and then use the hashtag {\it \#rogerfederer} for the next 5 days, while, community $C_2$ would use {\it \#federer} for the first 4 days and then {\it \#rogerfederer} for the next 3 days. The overall graph structure will suggest a majority usage of {\it \#federer} for the first 2 days (since both $C_1$ and $C_2$ use this hashtag in the first 2 days), a mixed usage for the next 2 days (since $C_1$ uses one and $C_2$ uses another hashtag during this period) and a majority usage of {\it \#rogerfederer} in the final 3 days. However, within the graph, the evolutions have a clear boundary - they are distinct, without much mixing, when investigated atomically from the standpoint of communities.
\end{itemize}

We demonstrate the effectiveness of our approach using around 20-30\% of eighteen days of Twitter data.
We observe that, all the four observations we have made, are novel in the literature.
Our work is the first of its kind in the space of Twitter topic lifecycle analysis, and presents insights that are fundamental for understanding the underlying dynamics of topics and their lifecycles.

\section{Related Work}
\label{sec:relwork}

The topic identification literature on Twitter has used three different approaches.
First, hashtags have been treated as topics, such as by \cite{cunha2011analyzing}.
Second, a burst of keywords in a short span of time are identified, and each bursting keyword is treated as a topic.
Works, such as \cite{cataldi2009cosena}, \cite{cataldi2010emerging} and \cite{mathioudakis2010twittermonitor}, use this.
Third, the latent semantic concepts of given tweets - often identified with sophisticated text-to-topic assignment techniques such as Latent Dirichlet Allocation (LDA) \cite{blei2003latent} - are treated as topics, and the tweets that address these spaces are said to belong to these topics.
Works, such as \cite{lau2012line}, follow this.
The first two approaches miss out on the latent semantic concept space addressed by the content, since they simply examine the keywords and hashtags instead of the overall content space.
Thus, these approaches would not be able to identify that tweets containing hashtags {\it \#mj}, {\it \#michaeljackson}, {\it \#jackson} and {\it \#m\_jackson} potentially address the same topic.
The third captures the semantic space of the concept inside the text well, but miss the explicit user intent expressed via hashtags.
Other works, such as \cite{ifrim2014event}, \cite{naaman2011hip} and \cite{narang2013discovery}, also use hashtag and LDA based methods for identifying topics, and analyzing their spatio-temporal evolution.

The Twitter topic lifecycle analysis literature has seen a strong work by Ardon {\it et al.} \cite{ardon2013spatio}.
They observe five phases in event lifecycles: pre-growth phase, growth phase, peak phase, decay phase and post-phase.
They perform the topic lifecycle analysis using individual hashtags as topics, and they further use a tool to identify places, entities {\it etc.} and assign these as tags (in turn, these tags become topics).
Amongst other works, the K-Spectral Centroid (KSC) clustering approach by Yang and Leskovec \cite{yang2011patterns} detects occurrence pattern similarity of hashtags, but does not consider any of, the time of occurrences, the semantic concept covered by the tweets having these hashtags, and the social network (friendship of users) aspects.
Stilo and Velardi \cite{stilo2017hashtag} propose SAX, that overcomes the temporal overlap aspect, but does not address the other two (semantic and social).

In general, the space of information diffusion has been extremely well-studied on Twitter.
Several works, such as Bakshy {\it et al.} \cite{bakshy2012role}, Kawk {\it et al.} \cite{kwak2010twitter} and Myers {\it et al.} \cite{myers2012information}, have investigated this problem.
Social affinity of discussions on Twitter has been observed by Narang {\it et al.} \cite{narang2013discovery}, and the geo-spatial characteristics of such discussions have been studied by Nagar {\it et al.} \cite{nagar2013topical}.
Many other works also galore.
An extensive survey of the literature, towards information diffusion and topic lifecycle analysis, has been conducted by Dey {\it et al.} \cite{dey2017literature}.

However, no work in the prior literature examines the lifecycle of a collection of hashtags with topics in the context of communities.
Further, none of the works attempt to investigate along the lines of correlating social communities with information topic lifecycles.
Our work, thus, is the first of its kind.
\section{Our Approach}
\label{sec:algo}

The input to our system is a collection of tweets that consisting of at least one hashtag.
The aim is to (a) create topics by creating clusters of semantically related hashtags with temporal overlap, (b) create communities, and (c) analyze the hashtag and topic lifecycles with respect to the communities, in terms of how topics morph over evolutions of hashtags within and across communities, as described in Section~\ref{sec:intro}.

The overview of our approach is as follows.
We create a timeline for the hashtags, tracking the usage frequency (count) of each hashtag within each timeslot.
We identify a word embedding for each hashtag using the content associated with it (since hashtags by themselves are non-dictionary words), using pre-trained embedding.
Using the similarity of embeddings as the distance measure for each pair of hashtags, we perform k-means clustering of the hashtags.
These clusters are further split such that, each hashtag present in a given (splitted) cluster temporally overlap in terms of occurrence.
Each cluster of hashtags (after splitting) is treated as a topic.
We identify modularity-based communities \cite{newman2006modularity} that are present in the underlying social network.
The hashtag usage of each user of a given community is aggregated to derive the hashtag usage made by the community, thereby creating a hashtag usage timeline of each community as a whole.
In addition, we overlap the topic cluster memberships of these hashtags, to create a topic participation timeline for each community as a whole.
These timelines are used to obtain hashtag-level and topic-level insights, in a community-agnostic manner as well as in the context of communities.

The details of our approach are provided below.

\subsection{Identifying ``Word Embedding'' of Hashtags}
\label{subsec:embedding}

We identify semantically related hashtags, using a word embedding technique followed by k-means clustering.\\

\noindent{\textbf{Step 1: Document creation}}\\
We create a document for each hashtag appearing in the dataset.
Let the set of hashtags present in the document be $H = \{h_1, h_2, h_3, ... \}$.
To this, we collect all the tweets $t_{h_i}$ where a given hashtag $h_i$ appears, and then append the tweets.
Thus for each hashtag $h_i$, we create a document $D_{h_i}$ as $D_{h_i} = \mathop{\bigcup}\{t_{h_i}\}$.\\

\noindent{\textbf{Step 2: Computing the ``word embedding'' of hashtags}}\\
In the next step, a word embedding model is created for each document (corresponding to a hashtag).
We eliminate all the hashtags occurring in document $D_{h_i}$, as well as, eliminate all the mentions.
We take the pre-trained Twitter-specific version of GloVe word embedding \cite{pennington2014glove} as an external resource, which has been learned on 2 billion tweets containing 27 billion tokens with a 1.2 million vocabulary size.
Let $W_{h_i}$ be the set of words appearing in $D_{h_i}$.
For each word $w_{h_i} \in W_{h_i}$, that is, each word that appears within the document of the hashtag, we look up the GloVe embedding of the word, and if found, we retain the word along with its embedding.
Finally, we compute an embedding $v_{h_i}$ for each given hashtag $h_i$ as a whole, using the embedding of the words that appear in the tweets containing the hashtag.
We compute this as the average of all the word embeddings that appear in its document.
\begin{equation}
v_{h_i} = \frac{\sum\limits_{w_{h_i} \in D_{h_i}}^{}(v_{w,h_i})}{|D_{h_i}|}
\label{eqn:Ehi}
\end{equation}

In Equation~\ref{eqn:Ehi}, $|D_{h_i}|$ represents the total length of the document $D_{h_i}$ as a count (total number) of the words appearing in the document, retaining words as many times as they appear.
The repeating behavior of words is retained, as this implicitly provides proportionate weight the embedding bears in the context of that hashtag: a word more used along with a given hashtag will get counted more frequently.
Further, in Equation~\ref{eqn:Ehi}, $v_{w,h_i}$ denotes the embedding of an individual word present in the pre-trained embedding.
The computation is repeated for all hashtags $h_i \in H$, creating a complete embedding map, for all the words $w_{h_i}$ under the context of all the hashtags $h_i$ that they appear in.

\subsection{Topic Cluster Creation using Related Hashtags}
\label{subsubsec:clustering}

\noindent{\textbf{Semantically related hashtag cluster creation}\\
We use the embeddings obtained in the earlier step, to obtain semantically related clusters.
In order to do this, we define a distance function for a given pair of embeddings: the value of cosine similarity of two given embeddings is treated as the distance between the pair of embeddings.
Cosine similarity of two vectors $v_1$ and $v_2$ (in this case, two embedding vectors) of dimension $d$ is given as:
\begin{equation}
similarity = cos(\theta) = \frac{\sum\limits_{i=1}^{d} v_{1_i}.v_{2_i}}{\sqrt{\sum\limits_{i=1}^{d} v_{1_i}^2}.\sqrt{\sum\limits_{i=1}^{d} v_{2_i}^2}}
\label{eqn:cosine}
\end{equation}

We now perform k-means clustering, in order to create clusters $T_s$ of conceptually (semantically) related hashtags.\\

\noindent{\textbf{Temporally relating hashtags for cluster creation}\\
Hashtags that would be contained in the same cluster, would be semantically as well as temporally related.
Hence, in the next step, we examine each semantically related cluster $t_s \in T_s$ in terms of temporal overlap.
Allen ~\cite{allen1983maintaining} created an exhaustive list of temporal relationships that can exist between a pair of time periods.
This includes {\it overlap}: partof event A and event B co-occur, {\it meets}: event A starts as soon as event B stops, and {\it disjoint}: event A and event B share no common time point.
In out setting, an event is an instance of a tweet using a given hashtag.

We create a time series of the individual hashtags, as well as the semantic clusters of hashtags obtained earlier.
For each timeslot, we compute whether or not a given hashtag is used.
We temporally relate a pair of hashtags $h_i$ and $h_j$ if they either satisfy the {\it overlaps} relationship, or if there exists one or more hashtags $h_k$, such that, $h_i$ is temporally related to $h_k$, and $h_k$ {\it overlaps} $h_j$, or, they are disjont by less than a threshold number of days (2 days for our experiments).
Two hashtags $h_i$ and $h_j$ are temporally unrelated if $\nexists h_k$ such that $h_i$ is temporally related to $h_k$, and, $h_k$ {\it overlaps} $h_j$.
The {\it temporally related} relationship is recursive in nature, and can be expressed as
\begin{equation}
h_i \odot h_j \implies \Big((\exists h_k) h_i \odot h_k\Big) \cap (h_k \circledcirc h_j)
\label{eqn:reltag}
\end{equation}
where $\odot$ denotes the {\it temporally related} relationship and $\circledcirc$ denotes the {\it overlaps} relationship.
A given semantic cluster $T_s$ will be split into two (or more) clusters $T_{s,t_1}$ and $T_{s,t_2}$, if there are two (or more) sets temporally related hashtags.\\

\noindent{\textbf{Topic cluster finalization}\\
We finalize our topics, defined as hashtag clusters, such that each hashtag cluster consists of hashtags that are both semantically and temporally related.
As an example, at the end of the process, hashtags such as $\{$\#tennis, \#federer, \#rogerfederer, \#roger$\}$ {\it etc.} are expected to be together in one cluster together if they occur closely in time, while hashtags such as $\{$\#politics, \#trump, \#donlandtrump, \#donald$\}$ {\it etc.} are expected to be together another cluster together.

\subsection{Creating Community-Level Hashtag and Topic Timelines}
\label{subsec:community}

Using the Twitter followership network of the users that posted the tweets, we discover modularity-based communities \cite{newman2006modularity}.
We subsequently perform aggregation of the users hashtag usage behavior, in order to find the total usage of each hashtag by community members, and find timelines.
Two timelines are found.

\noindent{\textbf{Hashtag-level usage timeline of communities}}\\
For each given timeslot, all the usages of a given hashtag for all the community members are summed up, to find the total number of usages of the hashtag by the community (that is, its members).
This gives the usage characteristics of each hashtag for each community, over each timeslot.
Further, we also note the topic cluster that each hashtag belongs to, which in turn gives, for each community, for each timeslot, a triplet
$$<community, timeslot, <topic\ and\ hashtag\ usage\ characteristics>>$$
wherein, each element within {\it $<$topic and hashtag usage characteristics$>$} consists of the following triplet
$$<hashtag, cluster\ of\ the\ hashtag, usage\ count\ of\ the\ hashtag>$$

\noindent{\textbf{Topic (cluster)-level usage timeline of communities}}\\
For each given timeslot, for each community, we sum up the usage count of all the hashtags belonging to the same topic cluster.
This is useful for identifying the participation of each given community in the topic as a whole, within the given timeslot.
This is captured in form of a triplet
$$<community, timeslot, <topic\ usage\ count\ over\ all\ hashtags>>$$
wherein, each element within {\it $<$topic usage count over all hashtags$>$} consists of the following pair
$$<cluster\ of\ the\ hashtag, usage\ count\ of\ all\ the\ hashtags\ in\ the\ cluster>$$

\subsection{Topic Lifecycle Analysis: Individual Topics and Communities}
\label{subsec:topiclifecycle}

We investigate two main aspects of topic lifecycles, both for our community-agnostic analysis as well as the analysis in the context of communities.\\

\noindent{\textbf{Dominant hashtag detection and topic morphing}}\\
A dominant hashtag is the one which has been most frequently used within a given timeslot, among all the hashtags.
In effect, it is the most representative hashtag of a topic at a given timeslot.
If a topic $t_k$ comprises of hashtags $H = \{_kh_1, _kh_2, ..., _kh_m\}$ for a given timeslot and if a function $g_c$ counts the number of times each hashtag $_kh_i$ was used, then, the dominant hashtag for the given timeslot is defined as
\begin{equation}
_kh_x = \forall(i) (max(g_c(_kh_i)))
\label{eq:max}
\end{equation}
While the traditional analysis of the dominant hashtag would tend to follow a lifecycle observed by Ardon {\it et al.} \cite{ardon2013spatio}, the lifecycle of the topic would be different, as over time, one dominant hashtag would take over another.
The change of the dominant hashtag of a given cluster over time, captures the morphing of the corresponding topic from being captured mostly by one hashtag to another.
The analysis is conducted at the level of communities also, in order to find the dominant hashtag usage made by each community at each timeslot and its evolution over time.
Note that, a topic morphs, when its dominant hashtag changes from one to the other.\\

\noindent{\textbf{Topic intensity detection}}\\
The intensity of a topic is derived as the summation of the number of times each hashtag is used.
We compute it both for the topics overall, as well as for each community.
It denotes the total presence of the topic (as a summation of the presence of its constituent hashtags) within the time slot, and in the other case, for each community.
If a topic $t_k$ comprises of hashtags $H = \{_kh_1, _kh_2, ..., _kh_m\}$ for a given timeslot and if a function $g_c$ counts the number of times each hashtag $_kh_i$ was used, then, the dominant hashtag for the given timeslot is defined as
\begin{equation}
_kh_x = \sum\limits_{i=1}^{m}(g_c(_kh_i))
\label{eq:sum}
\end{equation}
Note that, a topic dies, when its intensity becomes zero.
Further, if a topic intensity becomes zero within a community $C_1$ but is non-zero in another community $C_2$, it indicates that $C_1$ is no longer discussing the topic (the topic has died within community $C_1$) but $C_2$ is still discussing it (the topic is alive within $C_2$).


\section{Experiments}
\label{sec:expt}

\noindent{\textbf{Dataset Description}}\\
Our experiments use the tweet dataset\footnote{https://snap.stanford.edu/data/twitter7.html} by Yang and Leskovec \cite{yang2011patterns}.
It comprises of around 20\%-30\% of entire Twitter data of that period.
We use the data from $11^{th}$ to $30^{th}$ June 2009.
The corresponding social network connections data was obtained\footnote{http://an.kaist.ac.kr/traces/WWW2010.html} (Kwak {\it et al.} \cite{kwak2010twitter}).
We pre-process the data, to retain all the hashtags that occurred between $40$-$1,000$ times within this period.
This ensures that hashtags occurring frequently enough are retained, while the hashtags that associate with an excessively high number of tweets (mostly outliers) get ignored.
We retain the users that posted these tweets, and use the social connections among these users to form their social network subgraph.
The dataset is presented on Table~\ref{tab:datadesc}.

\begin{table}[htb]
\begin{center}
\begin{tabular}{|l|l|l|l|l|}
\hline
\textbf{Total num.} & \textbf{Num. hashtags} & \textbf{Num. tweets} & \textbf{Num. users} & \textbf{Avg. num. tweets} \\
\textbf{of tweets} & \textbf{retained} & \textbf{retained} & \textbf{retained} & \textbf{per user} \\
\hline
18,572,084 & 4,244 & 471,470 & 158,118 & 2.98 \\
\hline
\end{tabular}
\end{center}
\vspace{-0.2in}
\caption{Description of Available Data. All the tweets are from June 2009.}
\label{tab:datadesc}
\vspace{-0.3in}
\end{table}

\noindent{\textbf{Experimental Setup}}\\
We conduct our experiments on the given data, following the steps delineated in Section~\ref{sec:algo}.
We create 1-day timeslots for our experiments.
We use the BGLL algorithm \cite{blondel2008fast} for discovering communities.
We use the \textsc{kmeans} package of Python for doing k-means clustering.
We repeat our experiments at different granularities of k for finding clusters.
Since we have 4,244 hashtags, we range the value of k as $k = \{200, 400, 600, 800, 1000, 1200, 1400, 1600, 1800, 2000\}$, thus exploring at different clustering granularities.
We create the timeline for individual hashtags, for topics (clusters), and for the participation of communities in different topics over different hashtags across the different timeslots.\\

\noindent{\textbf{Inspecting the Topic Clusters}}\\
We examine the topic clusters derived by our process, to inspect the effectiveness of the embedding-and-clustering approach, given the relative novelty of this approach for clustering hashtags on Twitter data.
We present a few randomly chosen samples of topic clusters on Table~\ref{tab:clusters}.
Given space constraints, we have picked some of the k-values at random (k being the number of clusters in the corresponding k-means clustering), and have shown one randomly chosen topic cluster from each randomly chosen k-value.
It is visibly clear that the clusters are of consistently of good quality.\\

\begin{table}[htb]
\begin{center}
\begin{tabular}{|l|p{8.8cm}|}
\hline
\textbf{k-value} & \textbf{Cluster content} \\
\hline
2,000 & \#Nats, \#Rangers, \#WhiteSox \\
\hline
1,800 & \#musician, \#musiclover, \#singer \\
\hline
1,400 & \#Jackson, \#jackson, \#Rip, \#1984, \#jacko, \#kingofpop \\
\hline
1,000 & \#marijuana, \#drugwar, \#drugs, \#smoking \\
\hline
600 & \#Fashion, \#tshirts, \#shoes, \#makeup, \#clothing, \#sneakers, \#handbags \\
\hline
200 & \#cancer, \#Health, \#diet, \#medical, \#organic, \#weightloss, \#firstaid, \#ynw, \#healthy, \#nutrition, \#medicine, \#stemcells, \#Cancer, \#drugs, \#alcoholism, \#hiv, \#FDA \\
\hline
\end{tabular}
\end{center}
\vspace{-0.2in}
\caption{Examples of random clusters with random k-values (k of k-means clustering)}
\label{tab:clusters}
\vspace{-0.1in}
\end{table}

\begin{figure}[tbh]
\centering
\includegraphics[width=0.65\textwidth]{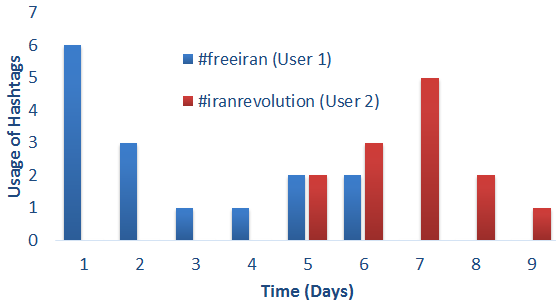}
\vspace{-0.1in}
\caption{Temporal overlaps of pairs of semantically related hashtags used by two random users}
\label{fig:userhashtags}
\vspace{-0.2in}
\end{figure}

\begin{figure}[tbh]
\centering
	\subfigure[Overall]{
		\includegraphics[width=0.3\textwidth]{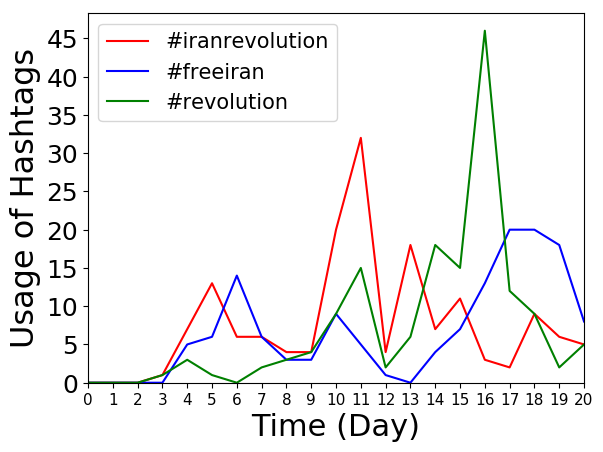}
		\label{fig:86.hashtag.timeseries}
	}
	\subfigure[Community 1]{
		\includegraphics[width=0.3\textwidth]{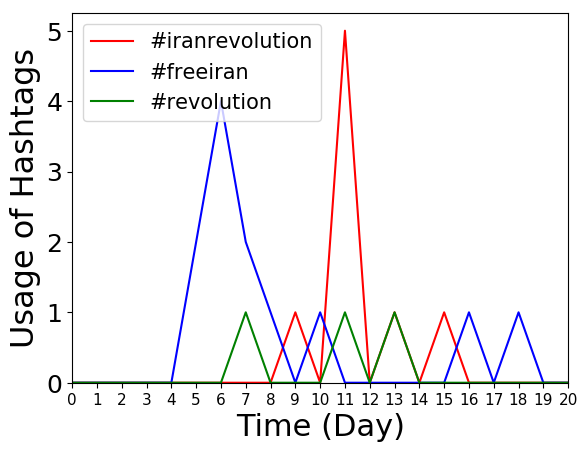}
		\label{fig:86.hashtag.timeseries.comm1}
	}
	\subfigure[Community 2]{
		\includegraphics[width=0.3\textwidth]{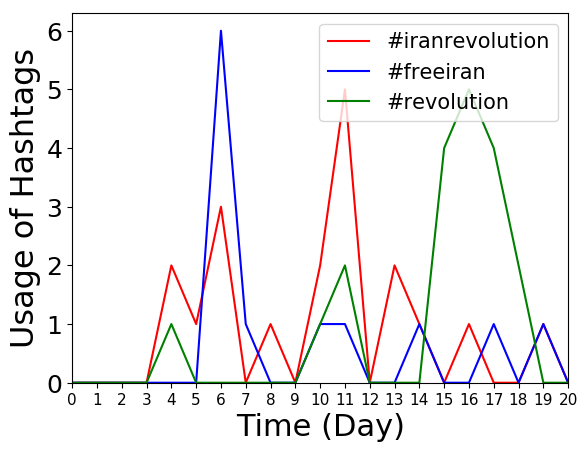}
		\label{fig:86.hashtag.timeseries.comm2}
	}
\vspace{-0.1in}
\caption{Time series of hashtags (iranrevolution, revolution, freeiran cluster)}
\label{fig:hashtag.iran}
\vspace{-0.4in}
\end{figure}

\begin{figure}[tbh]
\centering
	\subfigure[Overall]{
		\includegraphics[width=0.3\textwidth]{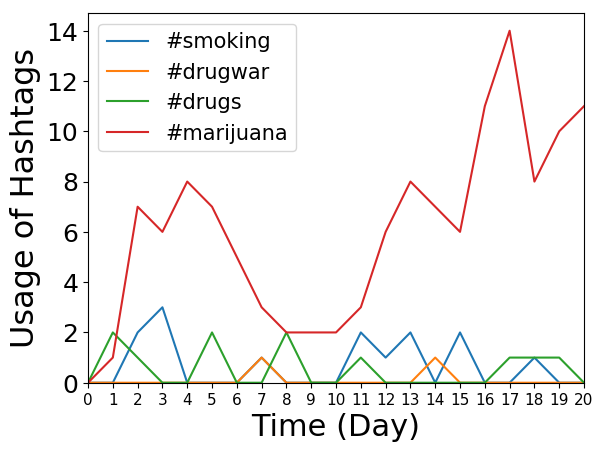}
		\label{fig:664.hashtag.timeseries}
	}
	\subfigure[Community 1]{
		\includegraphics[width=0.3\textwidth]{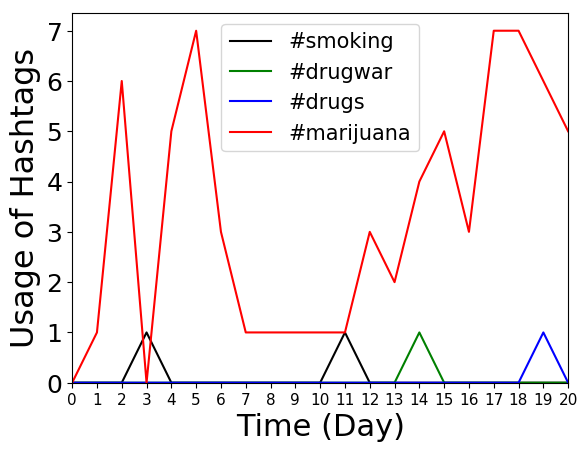}
		\label{fig:664.hashtag.timeseries.comm1}
	}
	\subfigure[Community 2]{
		\includegraphics[width=0.3\textwidth]{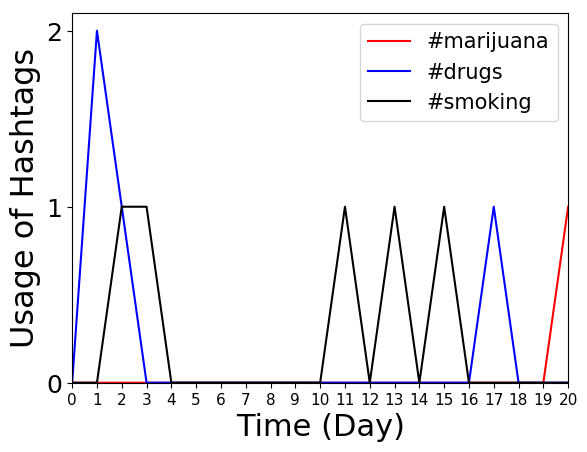}
		\label{fig:664.hashtag.timeseries.comm2}
	}
\vspace{-0.1in}
\caption{Time series of hashtags (drugs, smoking, drugwars, marijuana cluster)}
\label{fig:hashtag.drugs}
\vspace{-0.1in}
\end{figure}

\begin{figure}[tbh]
\centering
	\subfigure[Overall]{
		\includegraphics[width=0.3\textwidth]{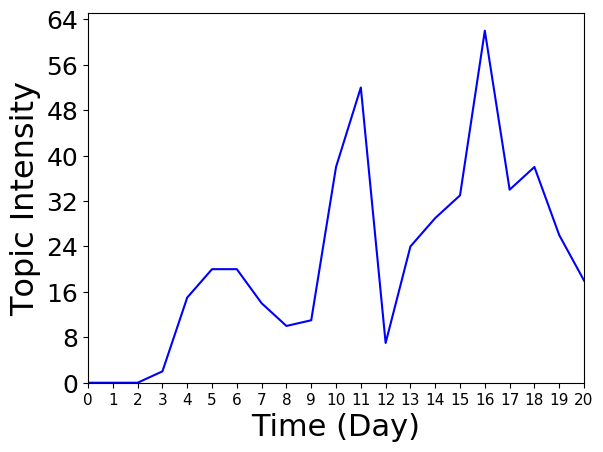}
		\label{fig:86.topic.timeseries}
	}
	\subfigure[Community 1]{
		\includegraphics[width=0.3\textwidth]{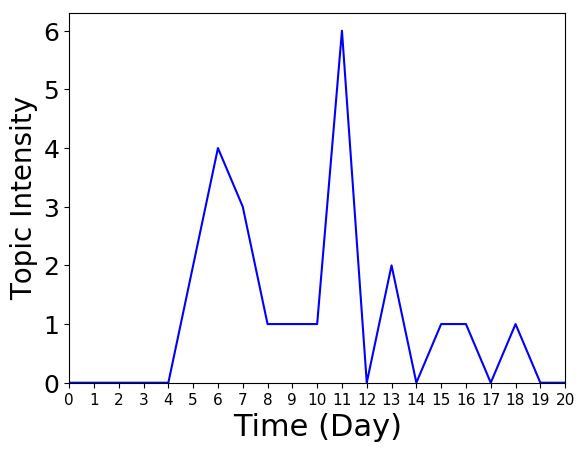}
		\label{fig:86.topic.timeseries.comm1}
	}
	\subfigure[Community 2]{
		\includegraphics[width=0.3\textwidth]{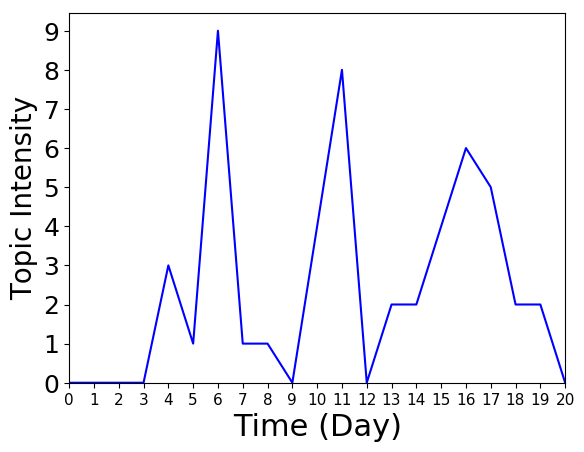}
		\label{fig:86.topic.timeseries.comm2}
	}
\vspace{-0.1in}
\caption{Time series of topic cluster (iranrevolution, revolution, freeiran cluster)}
\label{fig:topic.iran}
\vspace{-0.2in}
\end{figure}

\begin{figure}[tbh]
\vspace{-0.15in}
\centering
	\subfigure[Overall]{
		\includegraphics[width=0.3\textwidth]{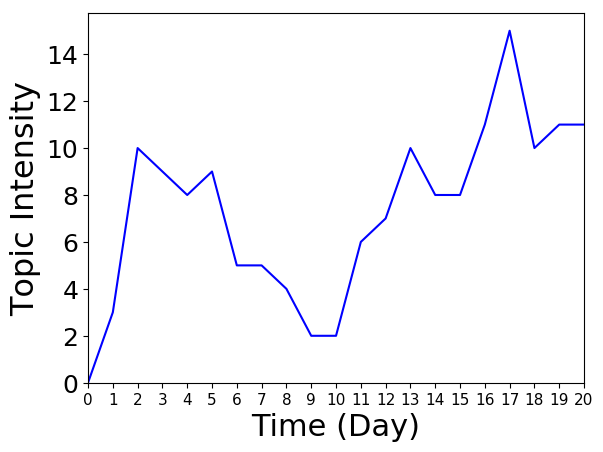}
		\label{fig:664.topic.timeseries}
	}
	\subfigure[Community 1]{
		\includegraphics[width=0.3\textwidth]{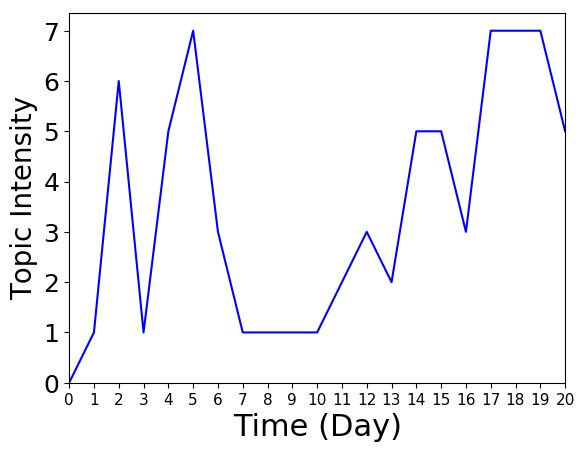}
		\label{fig:664.topic.timeseries.comm1}
	}
	\subfigure[Community 2]{
		\includegraphics[width=0.3\textwidth]{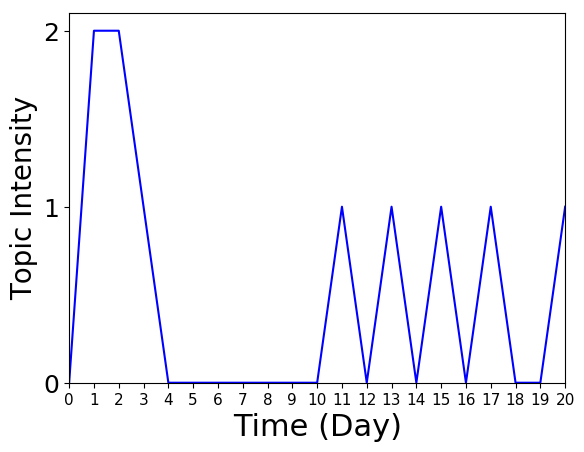}
		\label{fig:664.topic.timeseries.comm2}
	}
\vspace{-0.1in}
\caption{Time series of topic cluster (drugs, smoking, drugwars, marijuana cluster)}
\label{fig:topic.drugs}
\vspace{-0.4in}
\end{figure}

\noindent{\textbf{Topic Lifecycles - Overall and in Context of Communities: Our Findings}}\\
Our experiments provide strong support for all the four hypothesis we propose in our work.
We create the following kinds of plots to support our hypothesis.
\begin{enumerate}
\item {\it User participation plots:} These plots show the participation of given users to given hashtags (by virtue of the user using the hashtags).
\item {\it Hashtag lifecycle plots:} These plots show the overall lifespan of individual hashtags.
\item {\it Topic lifecycle plots:} These plots show the overall lifespan of the topics (clusters), aggregated across hashtags.
\item {\it Hashtag lifecycle plots per-community:} These plots show the lifespan of given individual hashtags, for a given community, indicating the participation of the community as a whole to these hashtags.
\item {\it Topic lifecycle plots per-community:} These plots show the overall lifespan of the topics (clusters), aggregated across hashtags, for a given community, indicating the participation of the community as a whole to a given topic.
\end{enumerate}

For qualitative analysis, we randomly choose two topic clusters from our dataset.
Cluster \textsc{C1} comprises of the hashtags {\it \#marijuana, \#drugwar, \#drugs, \#smoking} and cluster \textsc{C2} comprises of the hashtags {\it \#freeiran, \#iranrevolution, \#revolution}.
We randomly choose two users making sure that they are not connected with each other, and plot their hashtag usage characteristics towards cluster \textsc{C1} over time in Figure~\ref{fig:userhashtags}.
We observe that, they use different hashtags for the semantic concept captured by the cluster (one uses {\it \#freeiran} while the other uses {\it \#iranrevolution}).
On manual inspection, we see this behavior frequently repeating in the overall dataset, though we restrict to only one visual example here due to space constraints.
The observation supports our {\bf first hypothesis} - {\it conceptually related hashtags overlap semantically and temporally}.

We capture the timeseries of the individual hashtags in Figures \ref{fig:86.hashtag.timeseries} and \ref{fig:664.hashtag.timeseries}, and the timeseries of these hashtags with respect to two randomly chosen communities, respectively in Figures \ref{fig:86.hashtag.timeseries.comm1} and \ref{fig:86.hashtag.timeseries.comm2} for cluster \textsc{C1}, and Figures \ref{fig:664.hashtag.timeseries.comm1} and \ref{fig:664.hashtag.timeseries.comm2} for cluster \textsc{C2}.
It is visibly obvious from Figures \ref{fig:86.hashtag.timeseries.comm1} and \ref{fig:86.hashtag.timeseries.comm2} that, while the overall topic sees a good mix of all the hashtags (see Figure \ref{fig:86.hashtag.timeseries}), however, at given times, a given hashtag is clearly the dominant one in each community at a given time.
Since the hashtag usage at the level of a given community is simply the collective (aggregate) behavior of the members of the community, it entails that hashtag usage behavior is a community-level phenomenon.
This characteristic is reflected clearly in the other cluster as well.
These examples (and many others that we consistently observe, but do not report due to space constraints) corroborates our {\bf second hypothesis} - {\it hashtags associate with communities at a given time}, rather than independently among users.

Inspecting the community level hashtag usage timelines carefully, and comparing the hashtag usage behavior across the community pairs, the third and fourth hypothesis become clear.
For instance, comparing the hashtag usage behaviors shown in the figure pair Figure~\ref{fig:86.hashtag.timeseries.comm1} and \ref{fig:86.hashtag.timeseries.comm2}, it can be seen that although the hashtag {\it \#iranrevolution} follows similar dominance timelines across the two communities, the other hashtags have a different characteristics.
The hashtag {\it \#freeiran} is used from the $5^{th}$ to the $8^{th}$ day in \textsc{C1} but mostly from the $6^{th}$ to the $7^{th}$ day in \textsc{C2}.
Further, interestingly, the hashtag {\it \#revolution} remains absent in \textsc{C1} while strongly dominates in \textsc{C2}.
Such behavior is highly prominent in the figure pair Figure~\ref{fig:664.hashtag.timeseries.comm1} and \ref{fig:664.hashtag.timeseries.comm2}, where the hashtag {\it \#marijuana} is used in \textsc{C1} but practically not used in \textsc{C2}, while the hashtag {\it \#smoking} is used in \textsc{C2} but practically not used in \textsc{C1}.
All these collectively substantiate our {\bf third hypothesis} - {\it hashtags are independently used across communities}.
Further, the evolution of the hashtag {\it \#revolution} in \textsc{C1} acts as a demonstrative example of our {\bf fourth hypothesis} - {\it hashtags evolve independently (atomically) within communities}.
We also show the overall lifecycle of the corresponding topics, and their evolution, at an overall level in Figures \ref{fig:86.topic.timeseries} and \ref{fig:664.topic.timeseries}, and at a per-community level in Figures \ref{fig:86.hashtag.timeseries.comm1} and \ref{fig:86.hashtag.timeseries.comm2} for topic cluster \textsc{C1} and Figures \ref{fig:664.hashtag.timeseries.comm1} and \ref{fig:664.hashtag.timeseries.comm2} for topic cluster \textsc{C2}.

Note that, while we restrict our report to a small number of examples due to space constraints, we observe these characteristics to hold over a substantial volume of the data that we could manually inspect.

\section{Conclusion}
\label{sec:concl}

In this paper, we provided a novel analysis of topic lifecycles, in the context of social communities identified on Twitter.
We used semantically and temporally related clusters of hashtags as topics.
We used word embedding to enable hashtag clustering, thus ensuring the presence of higher order latent semantic space.
We provided novel insights on peculiarities of evolution of topics, manifested via usage of hashtags over time and the underlying social communities: hashtags (and topics) that remain within communities, topics that see the use of different hashtags in different communities at similar (overlapping) points of time, and topics that morph over hashtags within some communities while keep the hashtag used unchanged on other communities.
Empirically, we formed a baseline of hashtag lifecycles, and derived overall topic lifecycles by analyzing the aggregate characteristics of all hashtags in a given topic cluster.
Our experiments substantiated our set of hypotheses.
Our work would play a transformational role in the current understanding of information diffusion models, as well as, in understanding the social boundaries of topic lifecycles over time.

\bibliographystyle{splncs03}
\bibliography{bib}

\end{document}